\documentclass[twocolumn, aps, prb, amsmath, amssymb,floatfix]{revtex4-2}
\usepackage{graphicx,hyperref,float,mathtools}
\usepackage[version=4]{mhchem}
\usepackage{cleveref}
\usepackage{color}
\newcommand{\STO}{SrTiO$_\text{3}$}
\newcommand{\mycomment}[1]{}
\newcommand{\mub}{$\mu_{\rm B}$}

\begin{document}
\title{Disentangling the dynamics of transient spin and orbital magnetization in SrTiO$_3$ via the inverse Faraday effect from RT-TDDFT}

    \author{Andri Darmawan}
    \author{Markus E. Gruner}
    \author{Rossitza Pentcheva}

    \affiliation{Department of Physics and Center for Nanointegration (CENIDE), Universit\"at Duisburg-Essen, Lotharstr.~1, 47057 Duisburg, Germany}
 
	\date{\today{}}
\begin{abstract} 
Motivated by recent evidence of ferroelectricity and even multiferroicity in the prototypical diamagnetic band insulator SrTiO$_3$ from terahertz experiments, we investigate the carrier and magnetization dynamics of SrTiO$_3$ excited optically by linearly (LPL) and circularly polarized light (CPL). Our RT‑TDDFT simulations show a pronounced site‑ and orbital‑dependent charge transfer from O$2p$ to Ti $3d$ states. With LPL anti-phase oscillations of the electron‑density lobes at O and Ti resemble the soft transverse optical phonon mode and break dynamically inversion symmetry. CPL instead drives a coherent rotation of the charge dipoles around oxygen, producing a helicity‑dependent transient magnetization of opposite sign at O and Ti, even without ionic motion. 
The dominant effect stems from the transfer of angular momentum (AM) from light to the electronic orbital AM, while spin-orbit coupling plays a key role in the transfer from orbital to spin AM, the former being an order of magnitude larger than the latter. The real-time resolution allows us to disentangle the inverse Faraday effect, which follows the pulse envelope, from optical orientation, which builds up during the pulse and saturates afterwards. The results demonstrate that purely electronic processes without lattice motion induce ultrafast magnetisation in a non magnetic insulator, offering a tunable route for light controlled magnetic functionalities. 
\end{abstract}

\maketitle
Preparation and manipulation of transient magnetic states with light is a promising route towards novel applications, with ultrafast writing processes for magnetic storage technologies as a prominent example~\cite{Kimel2019,StanciuSwitching2007,Mangin2014,Lambert2014}.  The seminal work of Beaurepaire \emph{et al.} \cite{Beaurepaire1996} demonstrated the demagnetization of metallic ferromagnets by a single femtosecond laser pulse, paving the way for intensive reasearch on laser‐induced  phenomena in diverse classes of materials~\cite{Kirilyuk2010RMP,JuFMinduced2004,Kimel2004}.   In this context, real‑time time‑dependent density‑functional theory (RT‑TDDFT) has become a tool of choice for modeling a broad spectrum of light‑induced phenomena, including carrier dynamics in 2D materials~\cite{Li2021}, heterostructures~\cite{Gruner2019,Shomali2022,Shomali2024}, Weyl-semimetals~\cite{Sebesta2025} and non‑magnetic semiconductors~\cite{Bellersen2025} and has provided detailed insights into the redistribution of spin and orbital angular momentum during demagnetization~\cite{Krieger2015,DewhurstAngular2021,Mrudul2024demagnetizationFePt}.

The ultrafast demagnetization upon  optical absorption of linearly polarized light (LPL) is commonly related to thermal effects~\cite{Beaurepaire1996,Krieger2015,Mrudul2024demagnetizationFePt,mrudul2025generationphononsangularmomentum}. 
In contrast, circularly polarized light (CPL) involves a rotating vector potential that intrinsically breaks time‑reversal symmetry in the Hamiltonian.  The interaction of  CPL with materials allows for nonthermal excitations  via the  {\em inverse Faraday
effect} (IFE), where a time-dependent electric field induces an effective magnetic field~\cite{Pitaevskii1961}.
Originally introduced for transparent dielectrics in terms of classical electrodynamics~\cite{Pitaevskii1961,PershanIFE1966,vanderZielIFE1965}, material-specific quantum‑mechanical formulations~\cite{Popova2012,Battiato2014} revealed distinct spin‑ and orbital‑related contributions in metals that depend sensitively on material and photon energy~\cite{Berritta2016}.
IFE has been discussed for various materials including semimetals~\cite{Tokman-IFE_semimetals2020}, (non)magnetic metals~\cite{Berritta2016} or Mott insulators~\cite{Banerjee-IFEMott2022}.
An intriguing possibility is to induce magnetization  in nonmagnetic semiconducting or insulating materials, as recently reported for 2D materials, e.g. \ce{MoS2}~\cite{Okyay2020}, \ce{BiH}~\cite{Neufeld2023attosecond} or  oxides V$_2$O$_5$~\cite{Marini2022}. 

A related effect in semiconductors at frequencies above the bandgap, is {\em optical orientation} (OO)~\cite{zakharchenya1984optical}. It stems mainly from direct interband transitions between the valence and conduction band, whereas the IFE arises from nonlinear processes mediated by virtual electronic states. However, this conceptual distinction is not consistently applied to all classes of materials, specifically IFE is discussed also for metals, where direct interband transitions are expected to dominate~
\cite{Berritta2016,Mishra2023}. 

SrTiO$_3$ (STO) is a diamagnetic insulator~\cite{Onoda2011} and  a prototypical quantum paraelectric where quantum fluctuations suppress the ferroelectric (FE) transition~\cite{Mueller1979}.  It features superconducting upon doping~\cite{Schooley1964} and  serves as a versatile substrate for complex oxide electronics~\cite{Mannhart2010}, hosting a two‑dimensional electron gas at interfaces e.g. with LaAlO$_3$ \cite{Ohtomo2004} and also on its bare surface~\cite{SantanderSyro2011,Meevasana2011}.  The structural and electronic properties of bulk STO have been extensively characterized with DFT~\cite{Piskunov2004,Heifets2006,Wahl2008,El-Mellouhi2011}, while many‑body perturbation theory provides accurate optical and X‑ray absorption spectra~\cite{Sponza2013,Begum2019,Begum2023}.
Recent experiments have demonstrated that LPL with mid‑infrared~\cite{Nova2019} and THz~\cite{Li2019} frequencies can transiently induce ferroelectricity in STO, supported  by RT‑TDDFT simulations of THz~\cite{Shin2022} or optical excitations~\cite{Song-STO2023}. Beyond transient FE effects, 
Basini \emph{et al.} showed that CPL in the  THz range can induce multiferroicity in STO by exciting infrared phonon modes
~\cite{Basini2024terahertz}. The effect was attributed to the phonon analogue of the IFE -- the Barnett effect -- where coherent phonon motion induces a transient magnetization.  This dynamical multiferroicity~\cite{Juraschek2017} was linked to chiral phonons~\cite{Juraschek2025chiral} excited by CPL. Several theoretical approaches have been put forward to explain the  experimental observation~\cite{Basini2024terahertz} from different perspectives including second order perturbation theory with electron phonon coupling~\cite{Shabala-PhononIFE2024}, non-Maxwellian fields~\cite{Merlin2024PRB,Merlin2025}, electron-nuclear quantum geometry~\cite{Klebl2025} and atomic orbital magnetization~\cite{Urazhdin2025orbitalmomentgenerationcircularly}.
So far, the light-induced magnetization in STO has been limited to THz phonon excitation; to our knowledge, optically excited magnetisation by CPL has not been addressed so far.

In this work, we investigate the carrier and magnetization dynamics of STO under optical excitation with both linearly  and circularly polarized light using RT‑TDDFT.

By systematically varying the laser frequency and fluence we reveal a highly non‑trivial site‑ and orbital‑dependent carrier dynamics, with anti‑phase charge fluctuations at the oxygen and titanium sites for LPL reminiscent  of the soft transverse optical phonon mode. Most importantly, CPL in the optical frequency range generates a rotation of the charge dipoles around O and a sizable transient magnetic moment even without the prior excitation of phonon modes.  Our results shed light on the underlying mechanism with light helicity transferred to the electronic orbital angular momentum, while spin-orbit coupling turns out to be essential in inducing the substantially smaller spin contribution. 

To study the electronic and magnetic response of STO upon laser excitation, we performed RT-TDDFT calculations employing the Elk code~\cite{Elkcode} which implements the all-electron full-potential linearized augmented plane-wave (FP-LAPW) method. We used the generalized gradient approximation for the exchange correlation functional in the PBEsol implementation (GGA-PBEsol)~\cite{Perdew2008}. 
The non-collinear time-dependent Kohn-Sham equations are solved to obtain the time-dependent spinors 
\begin{equation} \label{eq.1}
    \begin{split}
    i &\frac{\partial \psi_j(\mathbf{r},t)}{\partial t} = \left[\frac{1}{2} \left(-i\nabla +\frac{1}{c}\mathbf{A}_{\mathrm{ext}}(t)\right)^2+v_s(\mathbf{r},t)\right. \\ 
    &+\frac{1}{2c}\boldsymbol{\sigma} \cdot \mathbf{B}_s(\mathbf{r},t)+ \left. \frac{1}{4c^2} \boldsymbol{\sigma}\cdot (\nabla v_s(\mathbf{r},t) \times( -i\nabla) \right] \psi_j(\mathbf{r},t),
\end{split}
\end{equation}
where $\boldsymbol{\sigma}$ is the vector of Pauli matrices. The effective Kohn-Sham (KS) potential $v_s(\mathbf{r},t) = v_{\mathrm{ext}}(\mathbf{r},t)+v_{\mathrm{H}}(\mathbf{r},t)+v_{xc}(\mathbf{r},t)$ consists of the external potential $v_{ext}$, the Hartree potential $v_{\mathrm{H}}$, and the exchange-correlation potential  $v_{\mathrm{xc}}$. The effective KS magnetic field $\mathbf{B}_s(\mathbf{r},t)=\mathbf{B}_{\mathrm{ext}}(\mathbf{r},t)+\mathbf{B}_{\mathrm{xc}}(\mathbf{r},t)$, contains the external magnetic field $\mathbf{B}_{\mathrm{ext}}$ and magnetic field stemming from the exchange-correlation potential $\mathbf{B}_{\mathrm{xc}}$. The final term in Eq.~\ref{eq.1} represents the spin-orbit coupling (SOC). We evaluate the total spin and orbital angular momentum by adding the contribution of each ion within the muffin tin (MT) sphere.
    \begin{figure}[tbh]
	   \hspace*{-1mm}
	   \includegraphics[width=0.8\columnwidth]{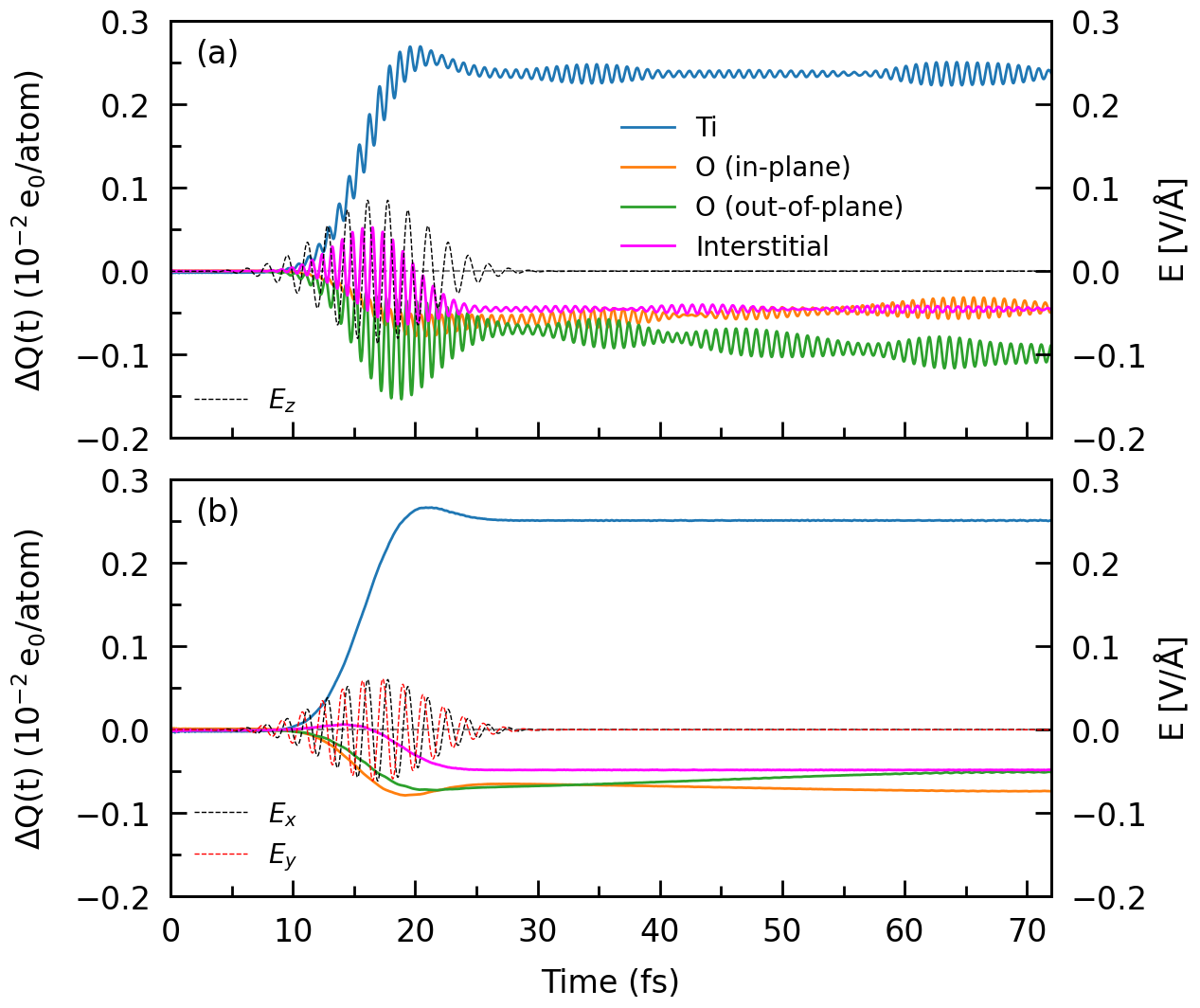}
	   \caption{Temporal evolution of the change in electronic charge $\Delta Q = Q(t)-Q(0)$ within the muffin-tin spheres of Ti, O and interstitial region following an excitation with (a) linearly and (b) circularly polarized light with a laser frequency of $\hbar\omega=2.5$~eV, laser peak intensity of $10^{11}$~W/cm$^2$  and a FWHM of 10 fs. The black dashed line depicts the $z$-component of the electric field of LPL, while black and red dashed lines represent the $x$- and $y$-components of the electric field of CPL.}
	\label{fig:DeltaQ-Lin_Circ}
    \end{figure}

The laser pulse  is described by a vector potential $\mathbf{A}_{\mathrm{ext}} (t)$, represented by a sinusoidal wave modulated by a Gaussian envelope function, within the dipole approximation. The electric field is obtained from the time-derivative of the vector potential, \(\textbf{E} = -\frac{1}{c} \frac{\partial \textbf{A}}{\partial t}\). The linearly polarized pulse is modeled as $\mathbf{A}_\mathrm{ext}(t) = \mathbf{A}_0 \exp{\left[-4\ln(2)\,\frac{(t-t_0)^2}{d^2}\right]}\sin\left[\omega(t-t_0)+\phi\right]$, where $\mathbf{A}_0$ is the vector polarization amplitude, $t_0$ is the laser peak time, $d$ is the full-width of half maximum (FWHM), $\omega$ is the laser frequency, and $\phi$ is the phase of the laser pulse. A circularly polarized pulse is modeled by a superposition of two linearly polarized pulses along $x$ and $y$ with a $\pm\pi/2$ phase shift.

     \begin{figure}[tbh]
       \centering
        \hspace*{-1mm}
        \includegraphics[width=\columnwidth,clip]{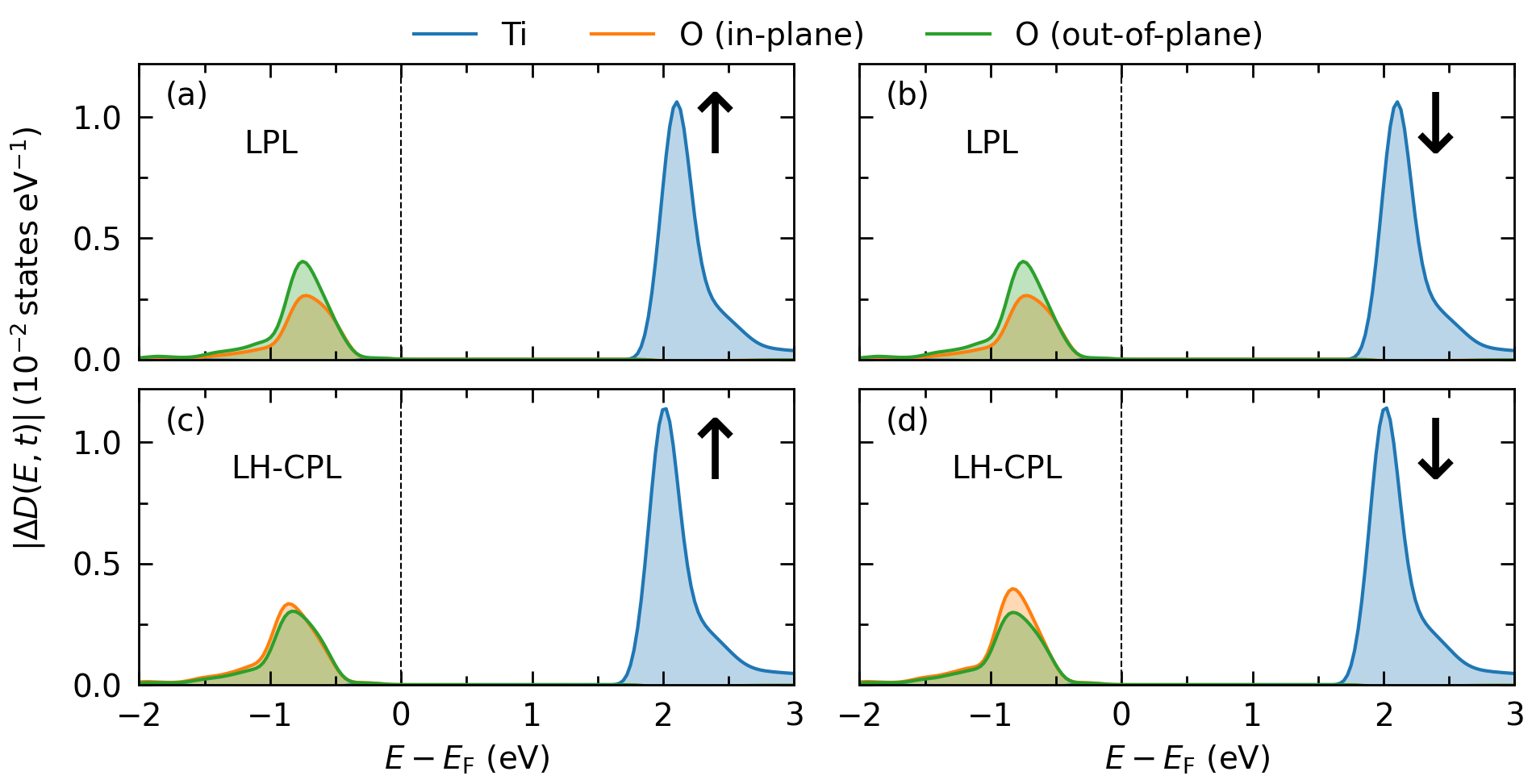}
        \caption{Changes of the spin- and element-resolved time-dependentdensity of states (DOS) during  the pulse ($t=17$~fs) for  (a-b) linearly and (c-d) circularly polarized light with laser frequency of $\hbar\omega=2.5$~eV with peak intensity $10^{11}$ W/cm$^2$. Changes in occupation at negative/positive energies denote initial/final states.} 
        \label{fig:TD-PDOS-LIN-vc-circ}
    \end{figure} 
    \begin{figure*}[tbh]
       \centering
       \hspace*{-1mm}
       \includegraphics[width=2\columnwidth,clip]{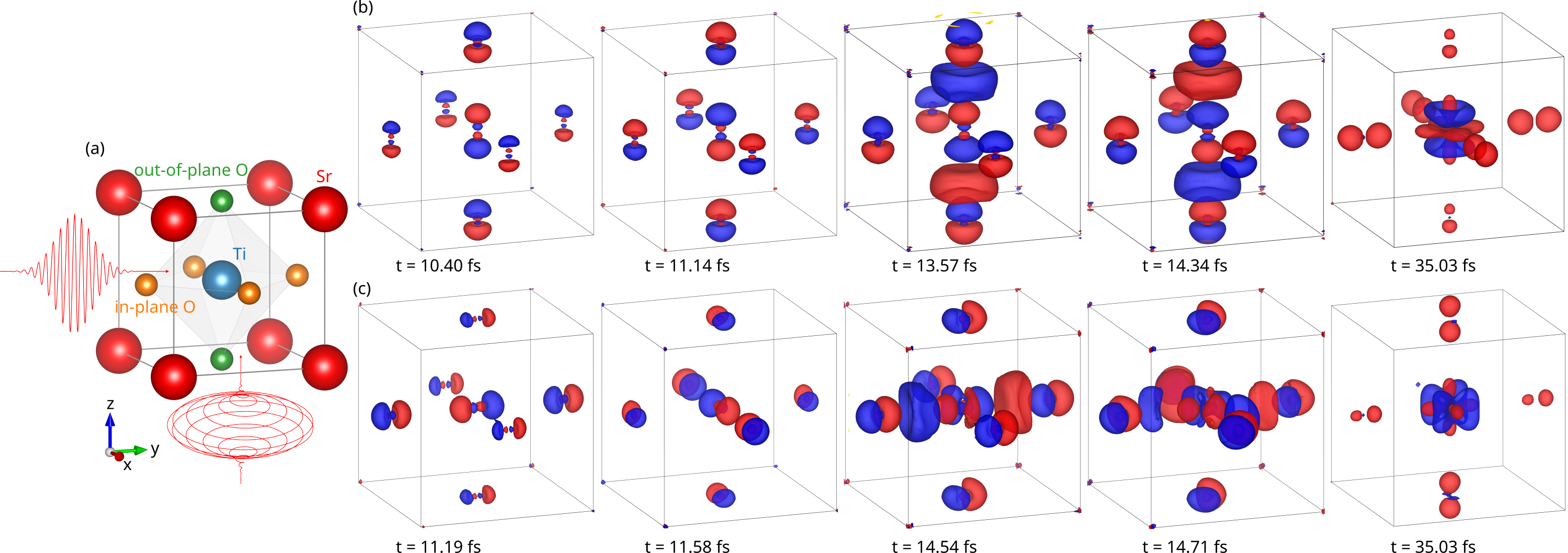}
       \caption{(a) Graphical illustration of linearly polarized light along the $z$-direction and circularly polarized light in the $xy$-plane driving the electronic excitation in \STO. Snapshots of the time-dependent electron density redistribution $\Delta \rho(\mathbf{r},t)$ during and after the laser excitation for (b) linearly and (c) circularly polarized pulse using laser frequency of $\hbar \omega= 2.5$~eV, FWHM = 10~fs, and laser peak intensity of S$_{\rm peak}$ = $10^{11}$ W/cm$^2$. Red and blue colors denote depletion and accumulation of charge, respectively. The isosurface level is $\pm$ 2.02 $\times 10^{-3} e_0/ \mathrm{\AA}^3$.} 
       \label{fig:RHO3D-LIN-vs-circ}
    \end{figure*}

\begin{figure}[bh]
	   \hspace*{-1mm}
	   \includegraphics[width=\columnwidth]{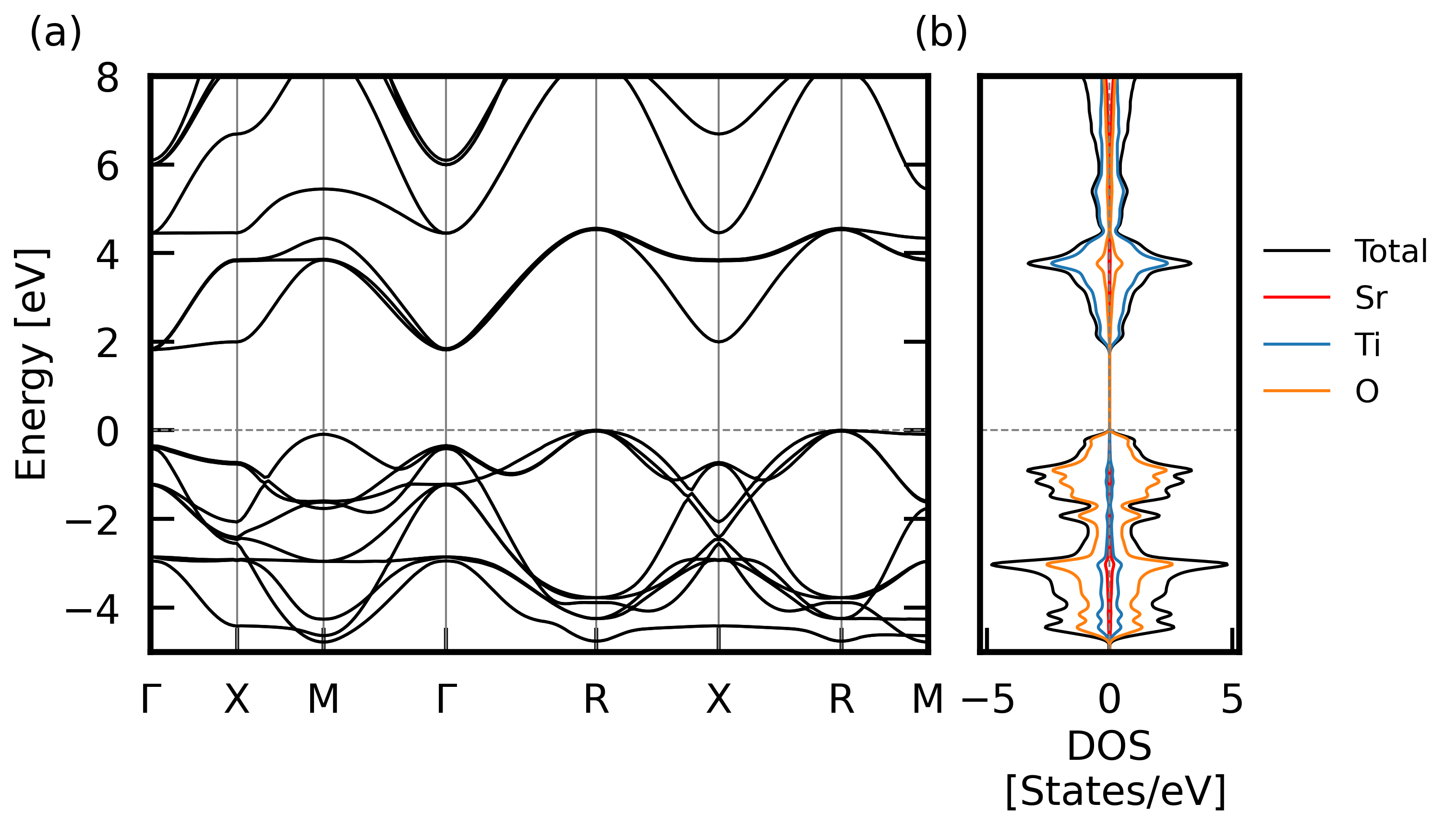}
	   \caption{(a) Electronic band structure along the high-symmetry directions and (b) projected density of states (PDOS) of \STO\ obtained with the  PBEsol exchange-correlation functional.}
	\label{fig:BandDOS}
    \end{figure}

In a first step we explore the charge dynamics of \STO\, upon laser excitation using linearly and circularly polarized light. The former has an electric field in $z$-direction, whereas the electric field vectors of left- (LH) and right-handed (RH) CPL lie in the $xy$-plane. We used a photon energy of 2.5~eV which is above the PBEsol band gap of 1.8~eV that separates occupied O $2p$ states from empty Ti $3d$ states, as shown in Fig.~\ref{fig:BandDOS}b. The laser pulse has a peak intensity of $10^{11}$~W/cm$^2$  with a FWHM of 10~fs which results in a  laser fluence of 0.38~mJ/cm$^2$. 

Fig.~\ref{fig:DeltaQ-Lin_Circ} shows the change in electronic charge within the MT spheres of O, Ti and the interstitial region for LPL and LH-CPL, whereas  no notable Sr-contribution is observed due to the absence of Sr states in the relevant energy region (see Fig.~\ref{fig:BandDOS}(b)). The most prominent feature is a depletion of charge at the O sites and the interstitial region and transfer to the Ti sites. The response to LPL and CPL is similar, except that oscillations in the CPL case are suppressed due to the superposition of two phase-shifted LPL pulses. 
Remarkably, the application of the light pulse persistently breaks the cubic symmetry of the \STO{} crystal lattice, lifting the degeneracy between in-plane and out-of-plane O sites, that exhibit a distinct dynamic response. In particular, for LPL a stronger depletion is observed at the out-of-plane oxygen sites. 

    The initial and final states involved in this charge transfer are unraveled by the time-dependent density of states $|\Delta D(E,t)|=|D(E,t)-D(E,0)|$, which displays the spin- and element-resolved (absolute) changes in the time dependent occupation of the KS orbitals with respect to the ground state DOS at $t=0$. Fig.~\ref{fig:TD-PDOS-LIN-vc-circ} shows $|\Delta D(E,t)|$ at $t=17$~fs, at the maximum of the pulse for LPL and CPL. We observe excitations from O~$2p$ states between $-1$ and $-0.5$~eV below $E_{\rm F}$ to empty  Ti~$3d$ states between $1.7$ and $2.3$~eV above $E_{\rm F}$, which is consistent with the laser frequency of $\hbar\omega=2.5$~eV. The depletion is distinct for in- and out-of-plane O $2p$ states, confirming the breaking of inversion symmetry. In contrast to LPL where the two spin channels show identical excitation, there is a noticeable difference between the majority and minority spin for CPL  in particular for the in-plane O $2p$ states.

A complementary and rather intuitive way to understand the charge dynamics induced by the laser excitation is offered by the electron density redistribution $\Delta\rho(\mathbf{r},t)=\rho(\mathbf{r},t)-\rho(\mathbf{r},0)$ with respect to the initial state. Fig.~\ref{fig:RHO3D-LIN-vs-circ}(b-c)  display snapshots with charge depletion (red) or accumulation (blue) primarily at the Ti and O sites in the first half ($t=10$ - $14$~fs) and after the pulse ($t=35\,$fs).

For LPL (Fig.~\ref{fig:RHO3D-LIN-vs-circ}(b)), oscillating positive and negative lobes form at the O and Ti sites along the field direction with opposite orientation at Ti and O sites, signifying a dynamic breaking of inversion symmetry. This dynamic pattern of the electronic clouds is reminiscent of the opposite  movement of Ti and O ions in the soft transverse optical TO$_1$ mode (e.\,g.,~\cite{Aschauer2014,kozina2019}. The soft TO mode hardens upon application of an electric field~\cite{Fleury1968,Akimov2000,Jacobsen2024} and may be considered as a signature of the incipient FE of SrTiO$_3$. An electronic excitation and  deformation of the electron density ~\cite{Song-STO2023} was previously related to the emergence of transient ferroelectricity~\cite{Li2019,Nova2019}

A remarkable feature for CPL (Fig.~\ref{fig:RHO3D-LIN-vs-circ}(c)) is the rotation of the lobes at the O sites  with the electric field, while the positive and negative lobes at the Ti site appear and disappear, predominantly oriented towards the in-plane oxygen ions, following the phase shift between the oscillating $x$- and $y$-components of the electric field. This establishes an effective circular current that is prone to transfer orbital angular momentum  from light to the electronic subsystem, as will be shown in the following. This circular motion of electron density is also expected to generate a circular motion of the ions.

After the decay of the laser pulse, the remaining charge redistribution concentrates at the Ti sites, whereas a predominant depletion of charge occurs at the in-plane O sites for LPL and at the out-of-plane oxygen for CPL. A complex shape emerges at the Ti sites with depletion from $e_g$ orbitals and accumulation into a linear combinations of out-of-plane $t_{2g}$ orbitals. 
    \begin{figure}[tbh]
      \centering
	   \hspace*{-1mm}
	   \includegraphics[width=\columnwidth]{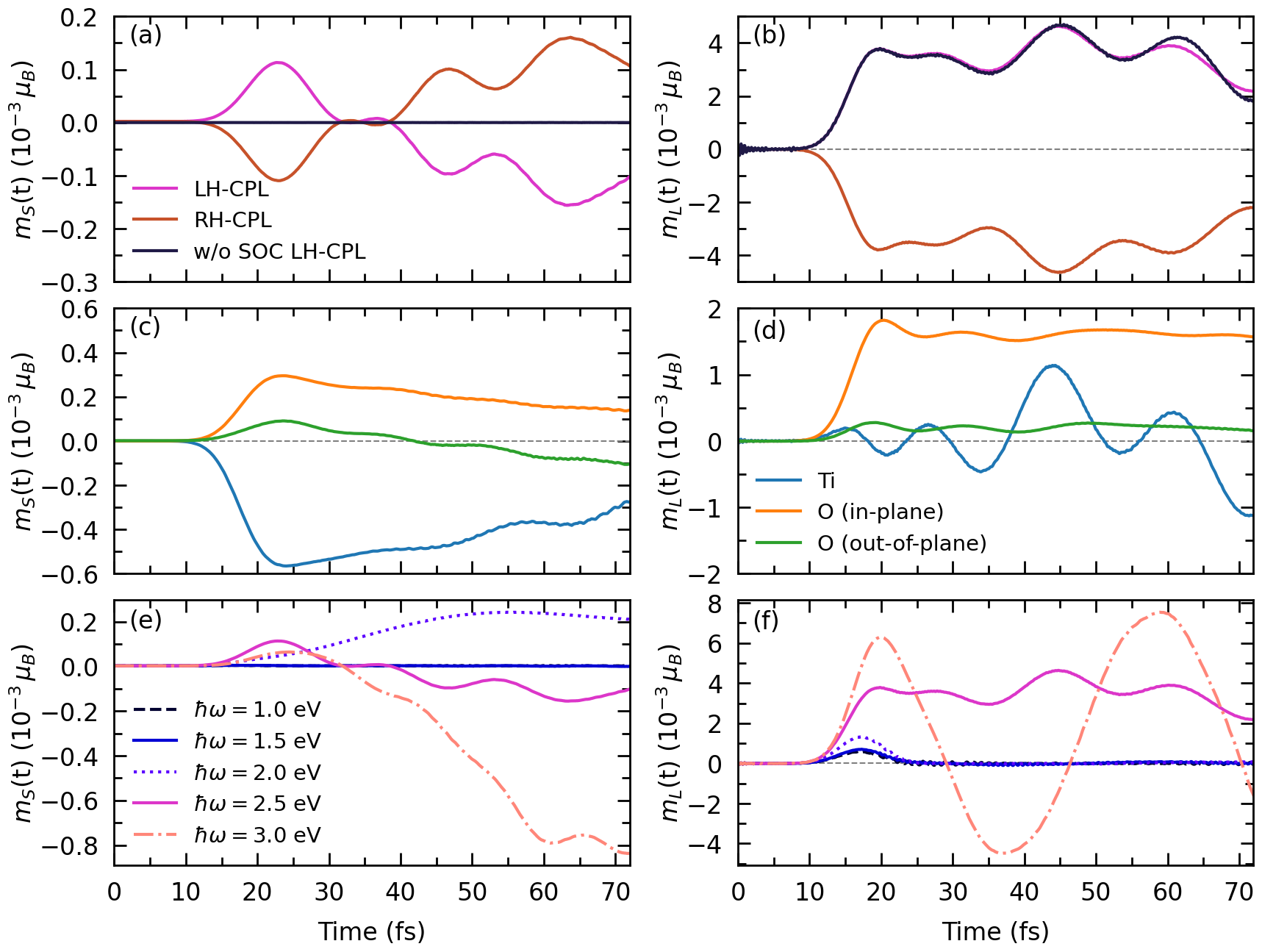}
	   \caption{Light-induced total (a) spin and (b) orbital magnetic moment for CPL with  $\hbar\omega$ = 2.5~eV. The absence of SOC, denoted by a light blue line, does not affect the orbital moment but quenches the spin moment. Element-resolved contributions to the laser-induced (c) spin and (d) orbital moment for LH-CPL. Comparison of the transient (e) spin  and (f) orbital magnetic moment induced by LH-CPL with different laser frequencies ranging from 1.0~eV to 3.0~eV, and S$_\mathrm{peak}$ = $10^{11}$ W/cm$^2$.  }
	\label{fig:mSmL-TotalandElements_circ_Helicity-SOC}
    \end{figure}
    
The circular motion of the positive and negative lobes of the electron density around the oxygen positions combined with the spin imbalance in the time-resolved DOS implies that CPL can induce a finite magnetization in the otherwise nonmagnetic \STO, which we elucidate in the following. Fig.~\ref{fig:mSmL-TotalandElements_circ_Helicity-SOC}(a) reveals  that a finite transient spin moment $m_{ S}(t)$ in the order of $10^{-4}\,\mu_{\rm B}$ develops during the pulse  for LH-CPL, which subsequently reverses direction reaching twice the magnitude at $t=70$~fs. For RH-CPL, $m_{S}(t)$ only changes its sign, confirming the helicity dependence. The oscillating behavior of $m_{S}(t)$ is a result of the delicate balance of larger, but competing element-specific contributions, as shown in Fig.~\ref{fig:mSmL-TotalandElements_circ_Helicity-SOC}(c): positive magnetization appears at the oxygen sites with a stronger contribution of in-plane oxygen and a pronounced negative contribution at the Ti sites. Initially, this yields a strong positive contribution from O which is not fully compensated by the antiparallel Ti moment.
After reaching a maximum toward the end of the pulse (25~fs), all element-specific contributions decrease in magnitude. The spin moment of out-of-plane oxygen eventually changes sign and the negative Ti moment dominates at larger times. Again, for RH-CPL, we obtain the same contributions with opposite sign (not shown), whereas no spin-polarization develops for LPL. 

Since the dipole selection rules usually prohibit a direct transfer of angular momentum from the photon to the electron spin, we elucidate the origin of the induced spin moment for CPL with $\hbar\omega=2.5$~eV and find that it  follows the occurrence of an orbital moment $m_{L}(t)$, which is one order of magnitude larger ($4\times 10^{-3}\,\mu_{\rm B}$), cf.~Fig.~\ref{fig:mSmL-TotalandElements_circ_Helicity-SOC}(b). $m_{L}(t)$ increases faster than $m_{S}(t)$ and reaches its first maximum approximately at the pulse maximum. After this, $m_{L}(t)$ preserves a constant magnitude overlaid by characteristic oscillations, originating mainly from the Ti sites, while the dominant contribution arises from in-plane oxygen, as shown in Fig.~\ref{fig:mSmL-TotalandElements_circ_Helicity-SOC}(d). RH-CPL, in turn, (not shown) mirrors exactly the trend with opposite sign. 

The evolution of $m_{S}(t)$ and $m_{L}(t)$ with the photon energy is displayed in Fig.~\ref{fig:mSmL-TotalandElements_circ_Helicity-SOC}(e,f). Laser excitations  below the GGA band gap  totally quench the spin moment due to the absence of electron excitation but still induce  a finite orbital moment solely during the pulse, which follows the Gaussian envelope of the pulse, reaching a maximum value of $0.9\times10^{-3}\,\mu_{\rm B}$. In contrast, photon energies above the gap yield a persistent 
 $m_{S}(t)$ and $m_{L}(t)$ even after the pulse. Their shape is, however, not uniform, depending strongly on frequency. For $\hbar\omega=2.0$~eV, $m_{L}(t)$ decays with the pulse, but induces a notable spin moment which evolves further in time.

For the optical excitation considered here, only the electrons can follow the rotation of the electric field and contribute to the evolution of transient magnetisation, while in the THz experiments of Basini et. {\em al.}~\cite{Basini2024terahertz} the frequency of the electric field oscillation is in the range of phonon modes. 
Urazhdin~\cite{Urazhdin2025orbitalmomentgenerationcircularly} proposed from a molecular model that the magnetization evolves from the time-dependent modification of hybridization between the orbitals of Ti and the surrounding O induced by the circularly polarized phonons. 
The electron density fluctuations in Fig.~\ref{fig:RHO3D-LIN-vs-circ} provide evidence for the orbital-dependent excitation, hybridization and  the rotation of induced dipoles in particular around the O sites driven by CPL in the optical frequency range.

A direct comparison to the experimentally measured magnetization ($1.6\times10^{-1}$ \mub)~\cite{Basini2024terahertz} 
is not straightforward, because the excitation energies differ: THz radiation in Ref.~\cite{Basini2024terahertz} versus optical excitation in the present work.

However, the magnetization predicted here from RT-TDDFT for CPL excitation in the optical range  ($m_L\sim 10^{-3}$~\mub) is two orders of magnitude higher than the one obtained from dynamical multiferroicity theory~\cite{Basini2024terahertz} ($\sim 10^{-5}$~\mub) and arises solely from the electronic degrees of freedom, thus signifying that a notable magnetization can also be induced with visible light.

Beyond the previous focus primarily on the transfer of  spin angular momentum ~\cite{Okyay2020,Marini2022,Neufeld2023attosecond}, we disentangle the mechanism of angular momentum transfer and the origin of both the spin and orbital contribution. The coincidence between the minima and maxima of oscillations in Fig.~\ref{fig:mSmL-TotalandElements_circ_Helicity-SOC}(a,b) indicates that spin- and orbital degrees of freedom are correlated, $m_{L}(t)$ being almost one order of magnitude larger than $m_{S}(t)$. While no significant spin-orbit coupling (SOC) is expected in \STO~\cite{Urazhdin2025orbitalmomentgenerationcircularly},  our results demonstrate the key role of SOC in the transfer of electronic orbital to spin angular momentum. In particular, switching off the spin-orbit coupling  completely quenches the spin moment, while the induced orbital moment is not affected (cf. Fig.~\ref{fig:mSmL-TotalandElements_circ_Helicity-SOC}a-b.) 

\begin{figure}[tbh]
	   \hspace*{-1mm}
	   \includegraphics[width=\columnwidth]{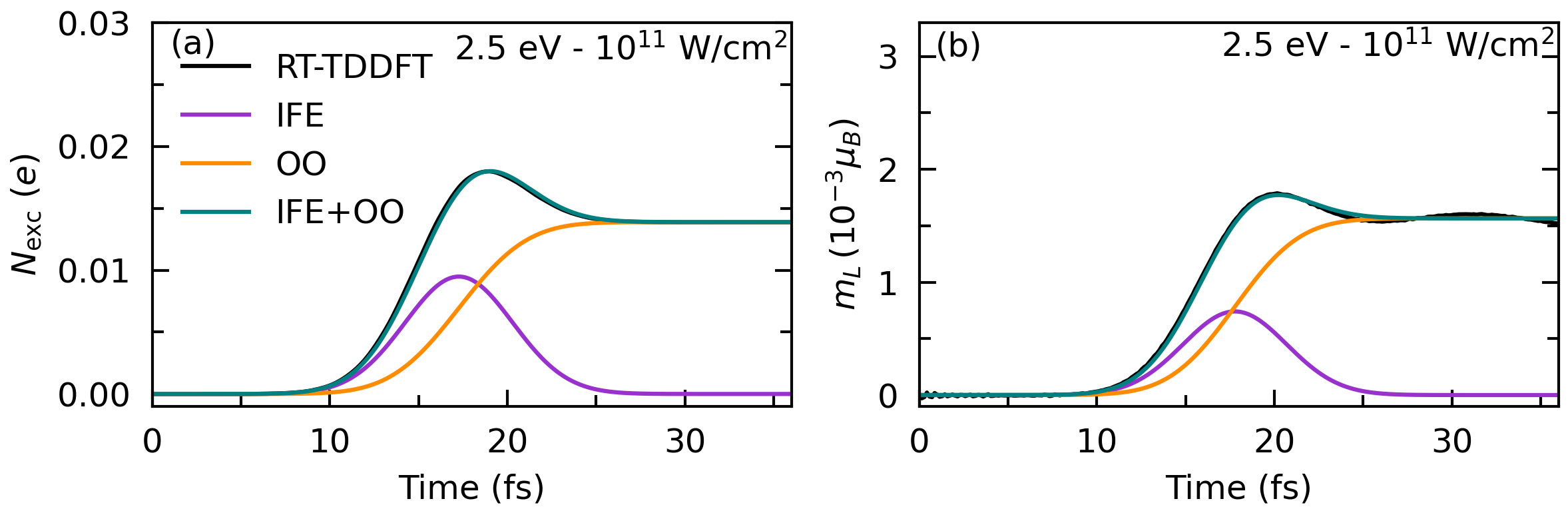}
	   \caption{(a) Transient number of excited electrons and (b) induced orbital magnetization at in-plane oxygen  decomposed into contributions from IFE and OO for a pulse with $\hbar\omega$ = 2.5~eV and laser intensity of $10^{11}$~W/cm$^2$.}
	\label{fig:IFE-vs-OO}
\end{figure}
\begin{figure}[bh]
   \hspace*{-1mm}
   \includegraphics[width=\columnwidth]{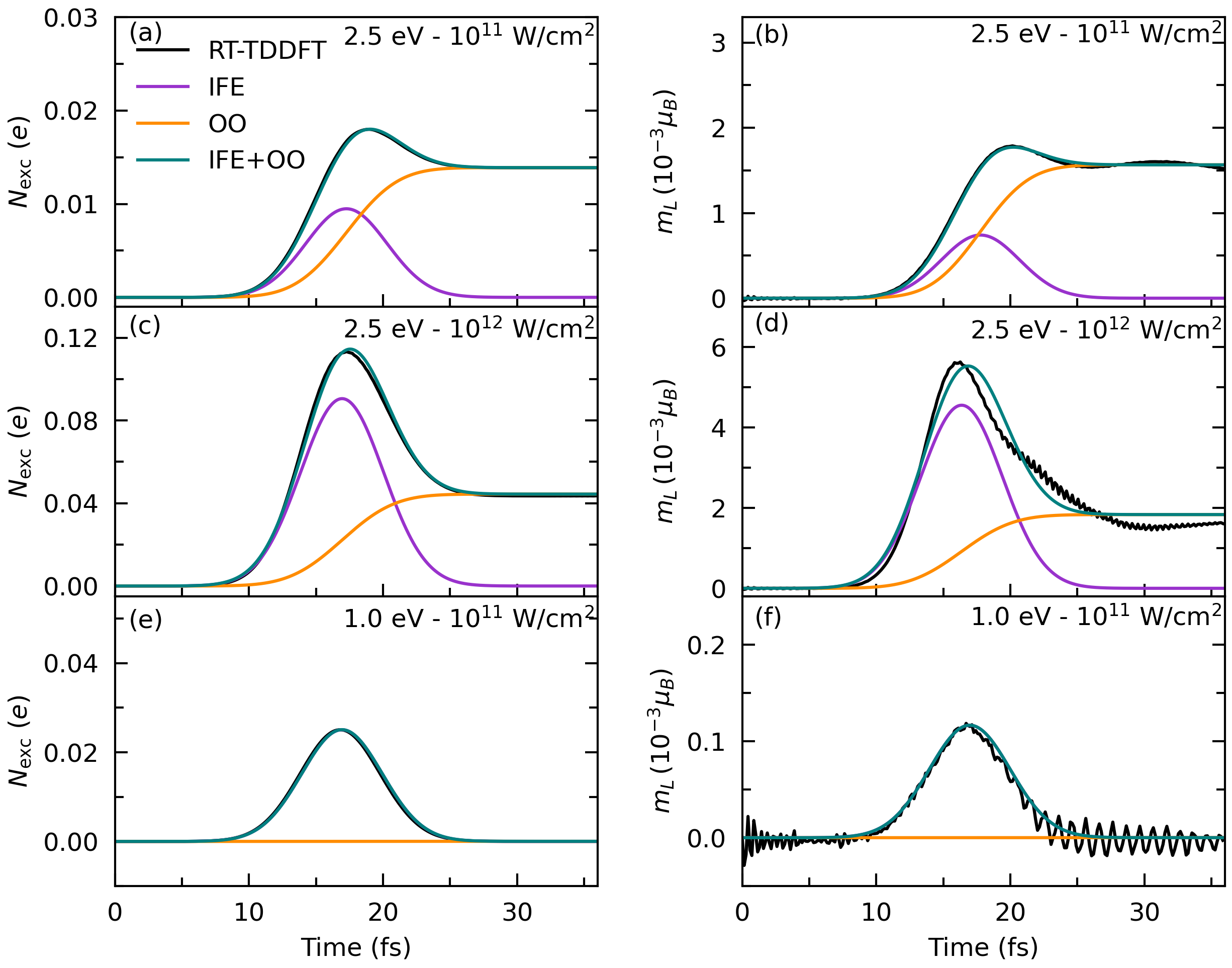}
   \caption{
     (a,c,e) Number of excited electrons and (b,d,f) induced orbital magnetization at the in‑plane O sites, modeled by 
     $%
       N_{\mathrm{exc}}(t)=N^{\mathrm{IFE}}\exp\!\left[-\frac{(t-t_{0})^{2}}{2\sigma^{2}}\right]
                           +\frac{N^{\mathrm{OO}}}{2}\Bigl[1+\operatorname{erf}\!
                            \bigl(\frac{t-t_{0}}{\sqrt{2}\,\sigma}\bigr)\Bigr]$
     and  
     $%
       m_{L}(t)=m_{L}^{\mathrm{IFE}}\exp\!\left[-\frac{(t-t_{0})^{2}}{2\sigma^{2}}\right]
               +\frac{m_{L}^{\mathrm{OO}}}{2}\Bigl[1+\operatorname{erf}\!
                \bigl(\frac{t-t_{0}}{\sqrt{2}\,\sigma}\bigr)\Bigr]$.
     Panels (a–b) correspond to an excitation above the gap
     ($\hbar\omega=2.5$~eV) with an intensity of $10^{11}$~W\,cm$^{-2}$;
     (c–d) show the same photon energy at $10^{12}$~W\,cm$^{-2}$; and
     (e–f) present excitation below the gap ($\hbar\omega=1.0$~eV) at
     $10^{11}$~W\,cm$^{-2}$.}
   \label{fig:Nexc-mL_IFE-vs-OO}
\end{figure}
\begin{figure}[tbh]
   \hspace*{-1mm}
   \includegraphics[width=\columnwidth]{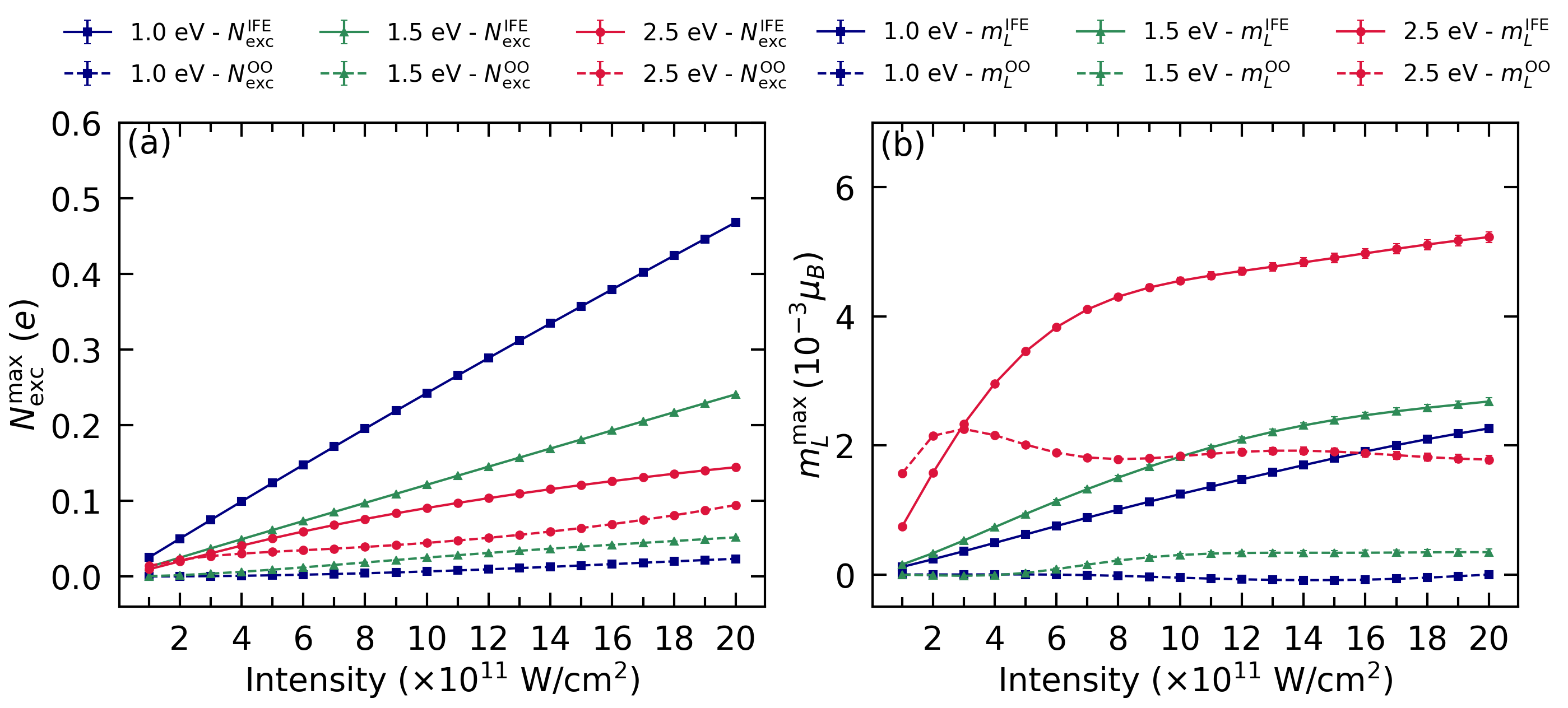}
   \caption{
     Maximum (a) number of excited electrons and (b) induced orbital
     magnetisation on the in‑plane O atoms for three photon energies
     (1.0, 1.5 and 2.5~eV), plotted as a function of laser intensity.
     The solid lines denote the IFE contribution, while the dashed lines
     represent the OO contribution.}
   \label{fig:Nexc-mL_IFE-vs-OO_intensity}
\end{figure}

Virtual transitions which give rise to the inverse Faraday effect are confined strictly to the duration of the light pulse and scale with the instantaneous intensity, which has a  Gaussian shape for the circularly‑polarised pulses applied here. In contrast, optical orientation contributions arise from direct carrier accumulation; they therefore grow proportionally to the transient pulse fluence (i.e., the time-integrated intensity) which for a Gaussian envelope can be expressed analytically by an error function.  
This allows us to decompose the IFE and OO contributions  in  Figure \ref{fig:IFE-vs-OO}  for   (a) the total number of excited carriers and (b) the transient orbital magnetization of the in‑plane O\,2$p$ states, which constitute the dominant initial‐state manifold. The results for additional frequencies are displayed in Figure \ref{fig:Nexc-mL_IFE-vs-OO}. A systematic deconvolution of IFE and OO as a function of pulse intensity (Fig.~\ref{fig:Nexc-mL_IFE-vs-OO_intensity}(a–b) in End Matter) reveals that, for moderately strong pulses, the induced magnetization for excitations below the band gap is solely due to the IFE. For photon energies above the gap, both mechanisms contribute, IFE mainly during the pulse, whereas OO develops with some delay and saturates after the pulse, with the nonlinear contribution becoming dominant at high intensities.

To summarize, our RT-TDDFT results demonstrate how ultrafast magnetisation can be optically induced in the non-magnetic band insulator \STO\ and uncover the circular motion of the orbitally polarized electronic charge clouds around the O ions as the mechanism of  angular momentum transfer of light  to the electronic orbital angular momentum.  
In turn, spin-orbit coupling is essential for  the transfer from orbital to spin angular momentum for above gap excitations.  Furthermore, we disentangle the contributions from two distinct mechanisms, IFE, dominating during the pulse even for subgap excitations and OO which emerges for frequencies above the band gap and saturates after the pulse.  Magnitude and temporal profile of the induced dynamics vary with laser parameters, indicating a tunable control over non-equilibrium states. Our work thus opens a new avenue for manipulating nonmagnetic materials   with time‑dependent electromagnetic fields to induce transient magnetization.

$Acknowledgments$---We gratefully acknowledge funding by the Deutsche Forschungsgemeinschaft (DFG, German Research Foundation) within collaborative research center CRC1242 (Project No.\ 278162697, subproject C02) and computational time at Leibniz
Rechenzentrum, project pr87ro and the Center for Computational Sciences and Simulation of the University of Duisburg-Essen on the supercomputer magnitUDE and amplitUDE (DFG Grants INST 20876/209-1 FUGG, INST 20876/243-1 FUGG and INST 20876/423-1 FUGG)

\bibliography{reference}

@article{Scali2026,
  title = {Optically induced orbital polarization in bulk \ce{Ge} and \ce{GaAs}},
  author = {Scali, F. and Finazzi, M. and Bottegoni, F. and Zucchetti, C.},
  journal = {Phys. Rev. Res.},
  volume = {8},
  issue = {2},
  pages = {023163},
  numpages = {5},
  year = {2026},
  month = {May},
  publisher = {American Physical Society},
  doi = {10.1103/c71s-44nt},
  url = {https://link.aps.org/doi/10.1103/c71s-44nt}
}

@article{DAlessandro2020,
  title = {Real-time modeling of optical orientation in \ce{GaAs}: Generation and decay of the degree of spin polarization},
  author = {D'Alessandro, M. and Sangalli, D.},
  journal = {Phys. Rev. B},
  volume = {102},
  issue = {10},
  pages = {104437},
  numpages = {7},
  year = {2020},
  month = {Sep},
  publisher = {American Physical Society},
  doi = {10.1103/PhysRevB.102.104437},
  url = {https://link.aps.org/doi/10.1103/PhysRevB.102.104437}
}

@book{zakharchenya1984optical,
  title={Optical Orientation},
  author={Meier, F and Zakharchenya, Boris Petrovitch},
  editor={Meier, F. and Zakharchenya,  Boris Petrovitch},
  year={1984},
  publisher={Elsevier, Amsterdam North-Holland}
}

@article{Pitaevskii1961,
  title = {Electric Forces in a Transparent Dispersive Medium},
  author = {Pitaevskii, L.},
  journal = {Sov. Phys. JETP},
  volume = {12},
  pages = {1008},
  numpages = {0},
  year = {1961},
  month = {May},
  url = {http://www.jetp.ras.ru/cgi-bin/e/index/e/12/5/p1008?a=list}
}

@article{vanderZielIFE1965,
  title = {Optically-Induced Magnetization Resulting from the Inverse {F}araday Effect},
  author = {van der Ziel, J. P. and Pershan, P. S. and Malmstrom, L. D.},
  journal = {Phys. Rev. Lett.},
  volume = {15},
  issue = {5},
  pages = {190--193},
  numpages = {0},
  year = {1965},
  month = {Aug},
  publisher = {American Physical Society},
  doi = {10.1103/PhysRevLett.15.190},
  url = {https://link.aps.org/doi/10.1103/PhysRevLett.15.190}
}

@article{PershanIFE1966,
  title = {Theoretical Discussion of the Inverse {F}araday Effect, Raman Scattering, and Related Phenomena},
  author = {Pershan, P. S. and van der Ziel, J. P. and Malmstrom, L. D.},
  journal = {Phys. Rev.},
  volume = {143},
  issue = {2},
  pages = {574--583},
  numpages = {0},
  year = {1966},
  month = {Mar},
  publisher = {American Physical Society},
  doi = {10.1103/PhysRev.143.574},
  url = {https://link.aps.org/doi/10.1103/PhysRev.143.574}
}

@article{JuFMinduced2004,
  title = {Ultrafast Generation of Ferromagnetic Order via a Laser-Induced Phase Transformation in \ce{FeRh} Thin Films},
  author = {Ju, Ganping and Hohlfeld, Julius and Bergman, Bastiaan and van de Veerdonk, Ren\'e J. M. and Mryasov, Oleg N. and Kim, Jai-Young and Wu, Xiaowei and Weller, Dieter and Koopmans, Bert},
  journal = {Phys. Rev. Lett.},
  volume = {93},
  issue = {19},
  pages = {197403},
  numpages = {4},
  year = {2004},
  month = {Nov},
  publisher = {American Physical Society},
  doi = {10.1103/PhysRevLett.93.197403},
  url = {https://link.aps.org/doi/10.1103/PhysRevLett.93.197403}
}

@article{StanciuSwitching2007,
  title = {All-Optical Magnetic Recording with Circularly Polarized Light},
  author = {Stanciu, C. D. and Hansteen, F. and Kimel, A. V. and Kirilyuk, A. and Tsukamoto, A. and Itoh, A. and Rasing, Th.},
  journal = {Phys. Rev. Lett.},
  volume = {99},
  issue = {4},
  pages = {047601},
  numpages = {4},
  year = {2007},
  month = {Jul},
  publisher = {American Physical Society},
  doi = {10.1103/PhysRevLett.99.047601},
  url = {https://link.aps.org/doi/10.1103/PhysRevLett.99.047601}
}

@article{Tokman-IFE_semimetals2020,
  title = {Inverse {F}araday effect in graphene and {W}eyl semimetals},
  author = {Tokman, I. D. and Chen, Qianfan and Shereshevsky, I. A. and Pozdnyakova, V. I. and Oladyshkin, Ivan and Tokman, Mikhail and Belyanin, Alexey},
  journal = {Phys. Rev. B},
  volume = {101},
  issue = {17},
  pages = {174429},
  numpages = {12},
  year = {2020},
  month = {May},
  publisher = {American Physical Society},
  doi = {10.1103/PhysRevB.101.174429},
  url = {https://link.aps.org/doi/10.1103/PhysRevB.101.174429}
}

@article{Banerjee-IFEMott2022,
  title = {Inverse {F}araday effect in Mott insulators},
  author = {Banerjee, Saikat and Kumar, Umesh and Lin, Shi-Zeng},
  journal = {Phys. Rev. B},
  volume = {105},
  issue = {18},
  pages = {L180414},
  numpages = {5},
  year = {2022},
  month = {May},
  publisher = {American Physical Society},
  doi = {10.1103/PhysRevB.105.L180414},
  url = {https://link.aps.org/doi/10.1103/PhysRevB.105.L180414}
}

@Article{Okyay2020,
  author    = {Okyay, Mahmut Sait and Kulahlioglu, Adem Halil and Kochan, Denis and Park, Noejung},
  journal   = {Phys. Rev. B},
  title     = {Resonant amplification of the inverse {F}araday effect magnetization dynamics of time reversal symmetric insulators},
  year      = {2020},
  issn      = {2469-9969},
  month     = sep,
  number    = {10},
  pages     = {104304},
  volume    = {102},
  doi       = {10.1103/physrevb.102.104304},
  publisher = {American Physical Society (APS)},
}

@article{Beaurepaire1996,
  title = {Ultrafast Spin Dynamics in Ferromagnetic Nickel},
  author = {Beaurepaire, E. and Merle, J.-C. and Daunois, A. and Bigot, J.-Y.},
  journal = {Phys. Rev. Lett.},
  volume = {76},
  issue = {22},
  pages = {4250--4253},
  numpages = {0},
  year = {1996},
  month = {May},
  publisher = {American Physical Society},
  doi = {10.1103/PhysRevLett.76.4250},
  url = {https://link.aps.org/doi/10.1103/PhysRevLett.76.4250}
}

@article{Berritta2016,
  title = {Ab Initio Theory of Coherent Laser-Induced Magnetization in Metals},
  author = {Berritta, Marco and Mondal, Ritwik and Carva, Karel and Oppeneer, Peter M.},
  journal = {Phys. Rev. Lett.},
  volume = {117},
  issue = {13},
  pages = {137203},
  numpages = {5},
  year = {2016},
  month = {Sep},
  publisher = {American Physical Society},
  doi = {10.1103/PhysRevLett.117.137203},
  url = {https://link.aps.org/doi/10.1103/PhysRevLett.117.137203}
}

@article{Mishra2023,
  title = {Spin contribution to the inverse {F}araday effect of nonmagnetic metals},
  author = {Mishra, Shashi B. and Coh, Sinisa},
  journal = {Phys. Rev. B},
  volume = {107},
  issue = {21},
  pages = {214432},
  numpages = {13},
  year = {2023},
  month = {Jun},
  publisher = {American Physical Society},
  doi = {10.1103/PhysRevB.107.214432},
  url = {https://link.aps.org/doi/10.1103/PhysRevB.107.214432}
}

@article{Urazhdin2025orbitalmomentgenerationcircularly,
  title = {Atomic and interatomic orbital magnetization induced in \ce{SrTiO3} by chiral phonons},
  author = {Urazhdin, Sergei},
  journal = {Phys. Rev. B},
  volume = {111},
  issue = {21},
  pages = {214435},
  numpages = {7},
  year = {2025},
  month = {Jun},
  publisher = {American Physical Society},
  doi = {10.1103/m73s-fxdp},
  url = {https://link.aps.org/doi/10.1103/m73s-fxdp}
}

@article{Shabala-PhononIFE2024,
  title = {Phonon Inverse {F}araday Effect from Electron-Phonon Coupling},
  author = {Shabala, Natalia and Geilhufe, R. Matthias},
  journal = {Phys. Rev. Lett.},
  volume = {133},
  issue = {26},
  pages = {266702},
  numpages = {6},
  year = {2024},
  month = {Dec},
  publisher = {American Physical Society},
  doi = {10.1103/PhysRevLett.133.266702},
  url = {https://link.aps.org/doi/10.1103/PhysRevLett.133.266702}
}

@article{Basini2024terahertz,
  title={Terahertz electric-field-driven dynamical multiferroicity in \ce{SrTiO3}},
  author={Basini, Martina and Pancaldi, Matteo and Wehinger, Bj{\"o}rn and Udina, Mattia and Unikandanunni, Vivek and Tadano, Terumasa and Hoffmann, Matthias C and Balatsky, Alexander V and Bonetti, Stefano},
  journal={Nature},
  volume={628},
  number={8008},
  pages={534--539},
  year={2024},
  publisher={Nature Publishing Group UK London},
  doi = {https://doi.org/10.1038/s41586-024-07175-9}
}

@article{Merlin2024PRB,
  title = {Unraveling the effect of circularly polarized light on reciprocal media: Breaking time reversal symmetry with non-Maxwellian magnetic-esque fields},
  author = {Merlin, R.},
  journal = {Phys. Rev. B},
  volume = {110},
  issue = {9},
  pages = {094312},
  numpages = {5},
  year = {2024},
  month = {Sep},
  publisher = {American Physical Society},
  doi = {10.1103/PhysRevB.110.094312},
  url = {https://link.aps.org/doi/10.1103/PhysRevB.110.094312}
}

@article{Merlin2025,
    author = {Merlin, R.},
    title = {Magnetophononics and the chiral phonon misnomer},
    journal = {PNAS Nexus},
    volume = {4},
    number = {1},
    pages = {pgaf002},
    year = {2025},
    month = {01},
    issn = {2752-6542},
    doi = {10.1093/pnasnexus/pgaf002},
    url = {https://doi.org/10.1093/pnasnexus/pgaf002},
}

@article{Klebl2025,
  title = {Ultrafast Pseudomagnetic Fields from Electron-Nuclear Quantum Geometry},
  author = {Klebl, Lennart and Schobert, Arne and Eckstein, Martin and Sangiovanni, Giorgio and Balatsky, Alexander V. and Wehling, Tim O.},
  journal = {Phys. Rev. Lett.},
  volume = {134},
  issue = {1},
  pages = {016705},
  numpages = {7},
  year = {2025},
  month = {Jan},
  publisher = {American Physical Society},
  doi = {10.1103/PhysRevLett.134.016705},
  url = {https://link.aps.org/doi/10.1103/PhysRevLett.134.016705}
}

@article{Song-STO2023,
author = {Song, Chenchen and Yang, Qing and Liu, Xinbao and Zhao, Hui and Zhang, Cui and Meng, Sheng},
title = {Electronic Origin of Laser-Induced Ferroelectricity in \ce{SrTiO3}},
journal = {J. Phys. Chem. Lett.},
volume = {14},
number = {2},
pages = {576-583},
year = {2023},
doi = {10.1021/acs.jpclett.2c03078},
URL = { https://doi.org/10.1021/acs.jpclett.2c03078},
}

@article{Shin2022,
  title = {Simulating Terahertz Field-Induced Ferroelectricity in Quantum Paraelectric \ce{SrTiO3}},
  author = {Shin, Dongbin and Latini, Simone and Sch\"afer, Christian and Sato, Shunsuke A. and Baldini, Edoardo and De Giovannini, Umberto and H\"ubener, Hannes and Rubio, Angel},
  journal = {Phys. Rev. Lett.},
  volume = {129},
  issue = {16},
  pages = {167401},
  numpages = {6},
  year = {2022},
  month = {Oct},
  publisher = {American Physical Society},
  doi = {10.1103/PhysRevLett.129.167401},
  url = {https://link.aps.org/doi/10.1103/PhysRevLett.129.167401}
}

@article{Mrudul2024demagnetizationFePt,
  title = {Ab initio investigation of laser-induced ultrafast demagnetization of \ce{L$1_{0}$} \ce{FePt}: Intensity dependence and importance of electron coherence},
  author = {Mrudul, M. S. and Oppeneer, Peter M.},
  journal = {Phys. Rev. B},
  volume = {109},
  issue = {14},
  pages = {144418},
  numpages = {12},
  year = {2024},
  month = {Apr},
  publisher = {American Physical Society},
  doi = {10.1103/PhysRevB.109.144418},
  url = {https://link.aps.org/doi/10.1103/PhysRevB.109.144418}
}

@article{Neufeld2023attosecond,
  title={Attosecond magnetization dynamics in non-magnetic materials driven by intense femtosecond lasers},
  author={Neufeld, Ofer and Tancogne-Dejean, Nicolas and De Giovannini, Umberto and H{\"u}bener, Hannes and Rubio, Angel},
  journal={npj Comput. Mater.},
  volume={9},
  number={1},
  pages={39},
  year={2023},
  publisher={Nature Publishing Group UK London},
  doi = {https://doi.org/10.1038/s41524-023-00997-7}
}

@article{dewhurst2018oistr,
  title={Laser-induced intersite spin transfer},
  author={Dewhurst, John Kay and Elliott, Peter and Shallcross, Sam and Gross, Eberhard KU and Sharma, Sangeeta},
  journal={Nano Lett.},
  volume={18},
  number={3},
  pages={1842--1848},
  year={2018},
  doi = {https://doi.org/10.1021/acs.nanolett.7b05118},
  publisher={ACS Publications}
}

@article{DewhurstAngular2021,
  title = {Angular momentum redistribution in laser-induced demagnetization},
  author = {Dewhurst, J. K. and Shallcross, S. and Elliott, P. and Eisebitt, S. and Schmising, C. v. Korff and Sharma, S.},
  journal = {Phys. Rev. B},
  volume = {104},
  issue = {5},
  pages = {054438},
  numpages = {5},
  year = {2021},
  month = {Aug},
  publisher = {American Physical Society},
  doi = {10.1103/PhysRevB.104.054438},
  url = {https://link.aps.org/doi/10.1103/PhysRevB.104.054438}
}

@article{Wahl2008,
  title = {\ce{SrTiO3} and \ce{BaTiO3} revisited using the projector augmented wave method: Performance of hybrid and semilocal functionals},
  author = {Wahl, Roman and Vogtenhuber, Doris and Kresse, Georg},
  journal = {Phys. Rev. B},
  volume = {78},
  issue = {10},
  pages = {104116},
  numpages = {11},
  year = {2008},
  month = {Sep},
  publisher = {American Physical Society},
  doi = {10.1103/PhysRevB.78.104116},
  url = {https://link.aps.org/doi/10.1103/PhysRevB.78.104116}
}

@article{Begum2019,
  title = {Role of the exchange-correlation functional on the structural, electronic, and optical properties of cubic and tetragonal \ce{SrTiO3} including many-body effects},
  author = {Begum, Vijaya and Gruner, Markus Ernst and Pentcheva, Rossitza},
  journal = {Phys. Rev. Mater.},
  volume = {3},
  issue = {6},
  pages = {065004},
  numpages = {11},
  year = {2019},
  month = {Jun},
  publisher = {American Physical Society},
  doi = {10.1103/PhysRevMaterials.3.065004},
  url = {https://link.aps.org/doi/10.1103/PhysRevMaterials.3.065004}
}

@article{Begum2023,
  title = {Nature of excitons in the \ce{Ti} \emph{L} and \ce{O} \emph{K} edges of x-ray absorption spectra in bulk \ce{SrTiO3} from a combined first principles and many-body theory approach},
  author = {Begum-Hudde, Vijaya and Lojewski, Tobias and Rothenbach, Nico and Eggert, Benedikt and Eschenlohr, Andrea and Ollefs, Katharina and Gruner, Markus E. and Pentcheva, Rossitza},
  journal = {Phys. Rev. Res.},
  volume = {5},
  issue = {1},
  pages = {013199},
  numpages = {12},
  year = {2023},
  month = {Mar},
  publisher = {American Physical Society},
  doi = {10.1103/PhysRevResearch.5.013199},
  url = {https://link.aps.org/doi/10.1103/PhysRevResearch.5.013199}
}

@article{Gruner2019,
  title = {Dynamics of optical excitations in a \ce{Fe/MgO(001)} heterostructure from time-dependent density functional theory},
  author = {Gruner, Markus Ernst and Pentcheva, Rossitza},
  journal = {Phys. Rev. B},
  volume = {99},
  issue = {19},
  pages = {195104},
  numpages = {12},
  year = {2019},
  month = {May},
  publisher = {American Physical Society},
  doi = {10.1103/PhysRevB.99.195104},
  url = {https://link.aps.org/doi/10.1103/PhysRevB.99.195104}
}

@article{Shomali2022,
  title = {Anisotropic carrier dynamics in a laser-excited \ce{Fe1/(MgO)3}(001) heterostructure from real-time time-dependent density functional theory},
  author = {Shomali, Elaheh and Gruner, Markus E. and Pentcheva, Rossitza},
  journal = {Phys. Rev. B},
  volume = {105},
  issue = {24},
  pages = {245103},
  numpages = {12},
  year = {2022},
  month = {Jun},
  publisher = {American Physical Society},
  doi = {10.1103/PhysRevB.105.245103},
  url = {https://link.aps.org/doi/10.1103/PhysRevB.105.245103}
}

@article{Shomali2024,
author = {Shomali, Elaheh and Gruner, Markus E. and Pentcheva, Rossitza},
title = {Concerted Mechanism of Carrier Dynamics in Laser-Excited \ce{Fe/(MgO)(001)} Heterostructures from Real-Time Time-Dependent DFT},
journal = {Adv. Theory Simul.},
volume = {7},
number = {1},
pages = {2300319},
doi = {https://doi.org/10.1002/adts.202300319},
url = {https://advanced.onlinelibrary.wiley.com/doi/abs/10.1002/adts.202300319},
year = {2024}
}

@article{Nova2019,
author = {T. F. Nova  and A. S. Disa  and M. Fechner  and A. Cavalleri },
title = {Metastable ferroelectricity in optically strained \ce{SrTiO3}},
journal = {Science},
volume = {364},
number = {6445},
pages = {1075-1079},
year = {2019},
doi = {10.1126/science.aaw4911}
}

@article{Li2019,
author = {Xian Li  and Tian Qiu  and Jiahao Zhang  and Edoardo Baldini  and Jian Lu  and Andrew M. Rappe  and Keith A. Nelson },
title = {Terahertz field-induced ferroelectricity in quantum paraelectric \ce{SrTiO3}},
journal = {Science},
volume = {364},
number = {6445},
pages = {1079-1082},
year = {2019},
doi = {10.1126/science.aaw4913}
}

@article{Krieger2015,
author = {Krieger, K. and Dewhurst, J. K. and Elliott, P. and Sharma, S. and Gross, E. K. U.},
title = {Laser-Induced Demagnetization at Ultrashort Time Scales: Predictions of TDDFT},
journal = {J. Chem. Theory Comput.},
volume = {11},
number = {10},
pages = {4870-4874},
year = {2015},
doi = {10.1021/acs.jctc.5b00621},
URL = {https://doi.org/10.1021/acs.jctc.5b00621},

}

@article{mrudul2025generationphononsangularmomentum,
  title = {Generation of phonons with angular momentum during ultrafast demagnetization},
  author = {Mrudul, M. S. and Wei\ss{}enhofer, Markus and Oppeneer, Peter M.},
  journal = {Phys. Rev. B},
  volume = {112},
  issue = {18},
  pages = {L180407},
  numpages = {7},
  year = {2025},
  month = {Nov},
  publisher = {American Physical Society},
  doi = {10.1103/nt8w-47hb},
  url = {https://link.aps.org/doi/10.1103/nt8w-47hb}
}

@article{Marini2022,
  title = {Theory of ultrafast magnetization of nonmagnetic semiconductors with localized conduction bands},
  author = {Marini, Giovanni and Calandra, Matteo},
  journal = {Phys. Rev. B},
  volume = {105},
  issue = {22},
  pages = {L220406},
  numpages = {6},
  year = {2022},
  month = {Jun},
  publisher = {American Physical Society},
  doi = {10.1103/PhysRevB.105.L220406},
  url = {https://link.aps.org/doi/10.1103/PhysRevB.105.L220406}
}

@article{Elkcode,
    title ={The {E}LK code},
    author = {Dewhurst, J. K. and Sharma, S.},
    journal={elk.sourceforge.io},
    url = {https://elk.sourceforge.io/},
    year = {2024}
}

@article{Sponza2013,
  title = {Role of localized electrons in electron-hole interaction: The case of \ce{SrTiO3}},
  author = {Sponza, Lorenzo and V\'eniard, Val\'erie and Sottile, Francesco and Giorgetti, Christine and Reining, Lucia},
  journal = {Phys. Rev. B},
  volume = {87},
  issue = {23},
  pages = {235102},
  numpages = {11},
  year = {2013},
  month = {Jun},
  publisher = {American Physical Society},
  doi = {10.1103/PhysRevB.87.235102},
  url = {https://link.aps.org/doi/10.1103/PhysRevB.87.235102}
}

@article{kozina2019,
author={Kozina, M.
and Fechner, M.
and Marsik, P.
and van Driel, T.
and Glownia, J. M.
and Bernhard, C.
and Radovic, M.
and Zhu, D.
and Bonetti, S.
and Staub, U.
and Hoffmann, M. C.},
title={Terahertz-driven phonon upconversion in \ce{SrTiO3}},
journal={Nat. Phys.},
year={2019},
month={Apr},
day={01},
volume={15},
number={4},
pages={387-392},
issn={1745-2481},
doi={10.1038/s41567-018-0408-1},
}

@article{Fleury1968,
  title = {Soft Phonon Modes and the 110$^O$ {K} Phase Transition in \ce{SrTiO3}},
  author = {Fleury, P. A. and Scott, J. F. and Worlock, J. M.},
  journal = {Phys. Rev. Lett.},
  volume = {21},
  issue = {1},
  pages = {16--19},
  numpages = {0},
  year = {1968},
  month = {Jul},
  publisher = {American Physical Society},
  doi = {10.1103/PhysRevLett.21.16}
}

@article{Akimov2000,
  title = {Electric-Field-Induced Soft-Mode Hardening in \ce{SrTiO3} Films},
  author = {Akimov, I. A. and Sirenko, A. A. and Clark, A. M. and Hao, J.-H. and Xi, X. X.},
  journal = {Phys. Rev. Lett.},
  volume = {84},
  issue = {20},
  pages = {4625--4628},
  numpages = {0},
  year = {2000},
  month = {May},
  publisher = {American Physical Society},
  doi = {10.1103/PhysRevLett.84.4625},
  url = {https://link.aps.org/doi/10.1103/PhysRevLett.84.4625}
}

@article{Jacobsen2024,
  title = {Phonon dispersion of quantum paraelectric \ce{SrTiO3} in electric fields},
  author = {Jacobsen, Henrik and Barthkowiak, Marek and Weber, Tobias and Stuhr, Uwe and Roessli, Bertrand and Niedermayer, Christof and Staub, Urs},
  journal = {Phys. Rev. B},
  volume = {110},
  issue = {5},
  pages = {054302},
  numpages = {8},
  year = {2024},
  month = {Aug},
  publisher = {American Physical Society},
  doi = {10.1103/PhysRevB.110.054302},
  url = {https://link.aps.org/doi/10.1103/PhysRevB.110.054302}
}

@article{Kirilyuk2010RMP,
  title = {Ultrafast optical manipulation of magnetic order},
  author = {Kirilyuk, Andrei and Kimel, Alexey V. and Rasing, Theo},
  journal = {Rev. Mod. Phys.},
  volume = {82},
  issue = {3},
  pages = {2731--2784},
  numpages = {0},
  year = {2010},
  month = {Sep},
  publisher = {American Physical Society},
  doi = {10.1103/RevModPhys.82.2731},
  url = {https://link.aps.org/doi/10.1103/RevModPhys.82.2731}
}

@article{Aschauer2014,
doi = {10.1088/0953-8984/26/12/122203},
url = {https://dx.doi.org/10.1088/0953-8984/26/12/122203},
year = {2014},
month = {mar},
publisher = {IOP Publishing},
volume = {26},
number = {12},
pages = {122203},
author = {Aschauer, Ulrich and Spaldin, Nicola A},
title = {Competition and cooperation between antiferrodistortive and ferroelectric instabilities in the model perovskite \ce{SrTiO3}},
journal = {J. Phys. Condens. Matter.},
}

@article{Battiato2014,
  title = {Quantum theory of the inverse {F}araday effect},
  author = {Battiato, M. and Barbalinardo, G. and Oppeneer, P. M.},
  journal = {Phys. Rev. B},
  volume = {89},
  issue = {1},
  pages = {014413},
  numpages = {9},
  year = {2014},
  month = {Jan},
  publisher = {American Physical Society},
  doi = {10.1103/PhysRevB.89.014413},
  url = {https://link.aps.org/doi/10.1103/PhysRevB.89.014413}
}

@article{El-Mellouhi2011,
  title = {Modeling of the cubic and antiferrodistortive phases of \ce{SrTiO3} with screened hybrid density functional theory},
  author = {El-Mellouhi, Fedwa and Brothers, Edward N. and Lucero, Melissa J. and Scuseria, Gustavo E.},
  journal = {Phys. Rev. B},
  volume = {84},
  issue = {11},
  pages = {115122},
  numpages = {12},
  year = {2011},
  month = {Sep},
  publisher = {American Physical Society},
  doi = {10.1103/PhysRevB.84.115122},
  url = {https://link.aps.org/doi/10.1103/PhysRevB.84.115122}
}

@article{Heifets2006,
doi = {10.1088/0953-8984/18/20/009},
url = {https://dx.doi.org/10.1088/0953-8984/18/20/009},
year = {2006},
month = {may},
publisher = {},
volume = {18},
number = {20},
pages = {4845},
author = {Heifets, E and Kotomin, E and Trepakov, V A},
title = {Calculations for antiferrodistortive phase of \ce{SrTiO3} perovskite: hybrid density functional study},
journal = {J. Phys. Condens. Matter.},
}

@article{Juraschek2017,
  title = {Dynamical multiferroicity},
  author = {Juraschek, Dominik M. and Fechner, Michael and Balatsky, Alexander V. and Spaldin, Nicola A.},
  journal = {Phys. Rev. Mater.},
  volume = {1},
  issue = {1},
  pages = {014401},
  numpages = {9},
  year = {2017},
  month = {Jun},
  publisher = {American Physical Society},
  doi = {10.1103/PhysRevMaterials.1.014401},
  url = {https://link.aps.org/doi/10.1103/PhysRevMaterials.1.014401}
}

@article{Kimel2004,
  title={Laser-induced ultrafast spin reorientation in the antiferromagnet \ce{TmFeO3}},
  author={Kimel, AV and Kirilyuk, Andrei and Tsvetkov, Artem and Pisarev, RV and Rasing, Th},
  journal={Nature},
  volume={429},
  number={6994},
  pages={850--853},
  year={2004},
  publisher={Nature Publishing Group UK London},
  doi={https://doi.org/10.1038/nature02659}
}

@article{Lambert2014,
  title={All-optical control of ferromagnetic thin films and nanostructures},
  author={Lambert, Charles-Henri and Mangin, St{\'e}phane and Varaprasad, BSD Ch S and Takahashi, YK and Hehn, Michel and Cinchetti, M and Malinowski, Gr{\'e}gory and Hono, K and Fainman, Y and Aeschlimann, M and others},
  journal={Science},
  volume={345},
  number={6202},
  pages={1337--1340},
  year={2014},
  publisher={American Association for the Advancement of Science},
  doi={https://doi.org/10.1126/science.1253493}
}

@article{Mangin2014,
  title={Engineered materials for all-optical helicity-dependent magnetic switching},
  author={Mangin, St{\'e}phane and Gottwald, M and Lambert, CH and Steil, D and Uhl{\'\i}{\v{r}}, V and Pang, L and Hehn, Michel and Alebrand, S and Cinchetti, M and Malinowski, Gr{\'e}gory and others},
  journal={Nat. Mater.},
  volume={13},
  number={3},
  pages={286--292},
  year={2014},
  publisher={Nature Publishing Group UK London},
  doi={https://doi.org/10.1038/nmat3864}
}

@article{Mueller1979,
  title = {\ce{SrTiO3}: An intrinsic quantum paraelectric below 4 {K}},
  author = {M\"uller, K. A. and Burkard, H.},
  journal = {Phys. Rev. B},
  volume = {19},
  issue = {7},
  pages = {3593--3602},
  numpages = {0},
  year = {1979},
  month = {Apr},
  publisher = {American Physical Society},
  doi = {10.1103/PhysRevB.19.3593},
  url = {https://link.aps.org/doi/10.1103/PhysRevB.19.3593}
}

@article{Piskunov2004,
title = {Bulk properties and electronic structure of \ce{SrTiO3}, \ce{BaTiO3}, \ce{PbTiO3} perovskites: an ab initio {HF/DFT} study},
journal={Comput. Mater. Sci.},
volume = {29},
number = {2},
pages = {165-178},
year = {2004},
issn = {0927-0256},
doi = {https://doi.org/10.1016/j.commatsci.2003.08.036},
url = {https://www.sciencedirect.com/science/article/pii/S0927025603001812},
author = {S Piskunov and E Heifets and R.I Eglitis and G Borstel},
}

@article{Popova2012,
  title = {Theoretical investigation of the inverse {F}araday effect via a stimulated Raman scattering process},
  author = {Popova, Daria and Bringer, Andreas and Bl\"ugel, Stefan},
  journal = {Phys. Rev. B},
  volume = {85},
  issue = {9},
  pages = {094419},
  numpages = {13},
  year = {2012},
  month = {Mar},
  publisher = {American Physical Society},
  doi = {10.1103/PhysRevB.85.094419},
  url = {https://link.aps.org/doi/10.1103/PhysRevB.85.094419}
}

@article{Bellersen2025,
  title = {Ultrafast photoexcitation of semiconducting photocathode materials},
  author = {Bellersen, Hilde and Guerrini, Michele and Cocchi, Caterina},
  journal = {Phys. Rev. B},
  volume = {112},
  issue = {2},
  pages = {024314},
  numpages = {11},
  year = {2025},
  month = {Jul},
  publisher = {American Physical Society},
  doi = {10.1103/449c-nn6y},
  url = {https://link.aps.org/doi/10.1103/449c-nn6y}
}

@article{Li2021,
  title = {Ab initio study of ultrafast charge dynamics in graphene},
  author = {Li, Q. Z. and Elliott, P. and Dewhurst, J. K. and Sharma, S. and Shallcross, S.},
  journal = {Phys. Rev. B},
  volume = {103},
  issue = {8},
  pages = {L081102},
  numpages = {6},
  year = {2021},
  month = {Feb},
  publisher = {American Physical Society},
  doi = {10.1103/PhysRevB.103.L081102},
  url = {https://link.aps.org/doi/10.1103/PhysRevB.103.L081102}
}

@article{Ohtomo2004,
  author    = {Ohtomo, Akira and Hwang, Harold Y.},
  title     = {A high-mobility electron gas at the \ce{LaAlO3/SrTiO3} heterointerface},
  journal   = {Nature},
  volume    = {427},
  number    = {6973},
  pages     = {423--426},
  year      = {2004},
  doi       = {10.1038/nature02308},
}

@article{SantanderSyro2011,
  author    = {Santander-Syro, A. F. and Copie, O. and Kondo, T. and Fortuna, F. and Pailh{\`e}s, S. and Weht, R. and Qiu, X. G. and Bertran, F. and Nicolaou, A. and Taleb-Ibrahimi, A. and Le F{\`e}vre, P. and Herranz, G. and Bibes, M. and Reyren, N. and Apertet, Y. and Lecoeur, P. and Barth{\'e}l{\'e}my, A. and Rozenberg, M. J.},
  title     = {Two-dimensional electron gas with universal subbands at the surface of \ce{SrTiO3}},
  journal   = {Nature},
  volume    = {469},
  number    = {7329},
  pages     = {189--193},
  year      = {2011},
  doi       = {10.1038/nature09720}
}

@article{Meevasana2011,
  author    = {Meevasana, W. and King, P. D. C. and He, R. H. and Mo, S.-K. and Hashimoto, M. and Tamai, A. and Songsiriritthigul, P. and Baumberger, F. and Shen, Z.-X.},
  title     = {Creation and control of a two-dimensional electron liquid at the bare \ce{SrTiO3} surface},
  journal   = {Nat. Mater.},
  volume    = {10},
  number    = {2},
  pages     = {114--118},
  year      = {2011},
  doi       = {10.1038/nmat2943}
}

@article{Mannhart2010,
author = {J. Mannhart  and D. G. Schlom },
title = {Oxide Interfaces—An Opportunity for Electronics},
journal = {Science},
volume = {327},
number = {5973},
pages = {1607-1611},
year = {2010},
doi = {10.1126/science.1181862},
URL = {https://www.science.org/doi/abs/10.1126/science.1181862},
}

@article{Onoda2011,
doi = {10.1088/0953-8984/23/4/045604},
url = {https://doi.org/10.1088/0953-8984/23/4/045604},
year = {2011},
volume = {23},
number = {4},
pages = {045604},
author = {Onoda, Masashige and Tsukahara, Shuichi},
title = {The upper limit of thermoelectric power factors in the metal–band-insulator
crossover of the perovskite-type oxygen deficient system
\mbox{SrTiO}$_{3-\delta/2}$},
journal = {J. Phys.: Condens. Matter},

}

@article{Perdew2008,
  title = {Restoring the Density-Gradient Expansion for Exchange in Solids and Surfaces},
  author = {Perdew, John P. and Ruzsinszky, Adrienn and Csonka, G\'abor I. and Vydrov, Oleg A. and Scuseria, Gustavo E. and Constantin, Lucian A. and Zhou, Xiaolan and Burke, Kieron},
  journal = {Phys. Rev. Lett.},
  volume = {100},
  issue = {13},
  pages = {136406},
  numpages = {4},
  year = {2008},
  month = {Apr},
  publisher = {American Physical Society},
  doi = {10.1103/PhysRevLett.100.136406},
  url = {https://link.aps.org/doi/10.1103/PhysRevLett.100.136406}
}

@article{Schooley1964,
  title = {Superconductivity in Semiconducting \ce{SrTiO3}},
  author = {Schooley, J. F. and Hosler, W. R. and Cohen, Marvin L.},
  journal = {Phys. Rev. Lett.},
  volume = {12},
  issue = {17},
  pages = {474--475},
  numpages = {0},
  year = {1964},
  month = {Apr},
  publisher = {American Physical Society},
  doi = {10.1103/PhysRevLett.12.474},
  url = {https://link.aps.org/doi/10.1103/PhysRevLett.12.474}
}

@article{Kimel2019,
	title = {Writing magnetic memory with ultrashort light pulses},
	volume = {4},
	copyright = {2019 Springer Nature Limited},
	issn = {2058-8437},
	url = {https://www.nature.com/articles/s41578-019-0086-3},
	doi = {10.1038/s41578-019-0086-3},
	number = {3},
	urldate = {2026-01-08},
	journal = {Nat. Rev. Mater.},
	author = {Kimel, Alexey V. and Li, Mo},
	month = mar,
	year = {2019},
	pages = {189--200},
}

@article{Juraschek2025chiral,
author={Juraschek, Dominik M.
and Geilhufe, R. Matthias
and Zhu, Hanyu
and Basini, Martina
and Baum, Peter
and Baydin, Andrey
and Chaudhary, Swati
and Fechner, Michael
and Flebus, Benedetta
and Grissonnanche, Gael
and Kirilyuk, Andrei I.
and Lemeshko, Mikhail
and Maehrlein, Sebastian F.
and Mignolet, Maxime
and Murakami, Shuichi
and Niu, Qian
and Nowak, Ulrich
and Romao, Carl P.
and Rostami, Habib
and Satoh, Takuya
and Spaldin, Nicola A.
and Ueda, Hiroki
and Zhang, Lifa},
title={Chiral phonons},
journal={Nat. Phys.},
year={2025},
month={Oct},
day={01},
volume={21},
number={10},
pages={1532-1540},
issn={1745-2481},
doi={10.1038/s41567-025-03001-9},
}

@article{Sebesta2025,
author={{\v{S}}ebesta, Jakub and Gr{\aa}n{\"a}s, Oscar},
title={Photo-induced manipulation and relaxation dynamics of {W}eyl-semimetals},
journal={npj Comput. Mater.},
year={2025},
month={Jul},
day={07},
volume={11},
number={1},
pages={219},
issn={2057-3960},
doi={10.1038/s41524-025-01708-0},
url={https://doi.org/10.1038/s41524-025-01708-0},
}

    \begin{center}
        \textbf{End Matter}
    \end{center}

    Ground state and time-dependent electronic properties were obtained using the five atom cubic primitive cell of \STO\ with the PBEsol lattice constant of 3.894~\AA~\cite{Begum2019}.  We chose MT radii of 1.376~\AA, 1.097~\AA, and 0.823~\AA\, for Sr, Ti, and O, respectively. The plane-wave cutoff parameter, $RK_\mathrm{max}$, was set to 8. We employed a \textbf{\textit{k}}-mesh of \(10 \times 10 \times 10\) to sample the Brillouin zone for both the ground-state and the time-dependent calculations. As shown in  Fig.~\ref{fig:BandDOS}(a) and consistent with previous work~\cite{Wahl2008,Begum2019}, \STO\ is a band insulator with the valence band maximum (VBM) located at $R$ and  the conduction band minimum (CBM) at $\Gamma$, resulting in an  indirect ($R-\Gamma$) and direct band gap ($\Gamma-\Gamma$) of 1.81~eV and 2.16~eV with the PBEsol exchange correlation functional, respectively. Moreover, the projected density of states in Fig.~\ref{fig:BandDOS}(b) shows that the valence band is mostly comprised of O 2$p$ states, whereas the conduction band minimum has Ti $t_{2g}$ character. Many body effects (GW+BSE)~\cite{Sponza2013,Begum2019} are necessary to obtain the experimental band gap of 3.2 eV, but the shape and sequence of relevant bands is preserved.

    For the time-propagation a time-step size of $\Delta t = 0.726$~attoseconds was selected. Since phonons are not expected to be excited at the  photon energies in the optical range considered here, we concentrate solely on the electronic effects. 

The time-dependent density of states at a particular time $t$, $D(E,t)$, is calculated using~\cite{dewhurst2018oistr}:
\begin{equation}
    D(E,t)=\sum_{i=1}^\infty \int_{BZ} \delta(E-\epsilon_{i\mathbf{k}})g_{i\mathbf{k}}(t)
\end{equation}
with 
\begin{equation}
   g_{i\mathbf{k}}(t)=\sum_j n_{j\mathbf{k}} \int d^3r \psi^*_{i\mathbf{k}}(\mathbf{r},0)\psi_{j\mathbf{k}}(\mathbf{r},t)
\end{equation}
where $n_{j\mathbf{k}}$ is the occupation number of the $j^{th}$ orbital and $\psi_{i\mathbf{k}}(\mathbf{r},0)$ is the ground state Kohn-Sham orbital.

    \begin{figure}[tbh]
	   \hspace*{-1mm}
	   \includegraphics[width=0.8\columnwidth]{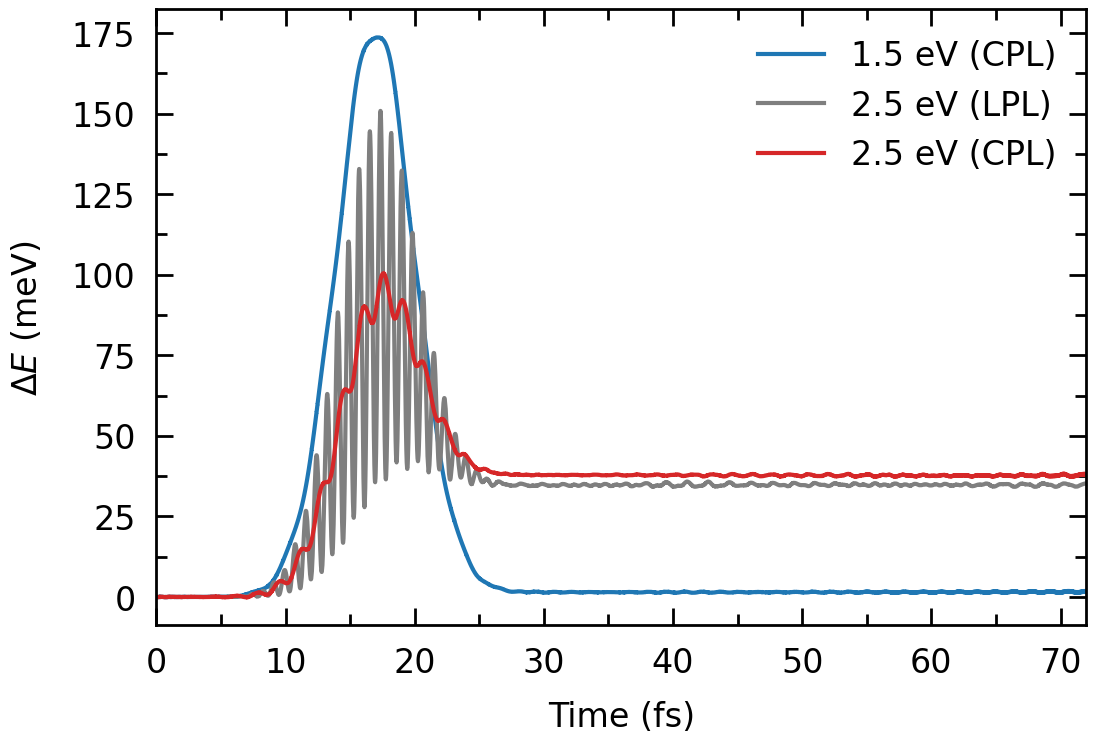}
	   \caption{The time evolution of changes in the total energy of the electronic system upon laser excitation for LPL and CPL with frequency of $\hbar\omega$ = 2.5~eV and CPL with laser frequency below the band gap ($\hbar\omega=1.5$~eV) with constant FWHM = 10~fs and laser peak intensity S$_\mathrm{peak}$ = $10^{11}$ W/cm$^2$}
	\label{fig:mSmL-TotalEnergy}
    \end{figure}

Changes in the time-dependent total Kohn-Sham energy during and after the pulse give insight into the material's response to the penetrating light indicating transient and permanent changes to the state of the material. Our simulations (cf. Fig.~\ref{fig:mSmL-TotalEnergy}) show a persistent energy uptake of $38$~meV/f.u. after the pulse for both LPL and CPL with $\hbar\omega=2.5$~eV beyond the GGA band gap, while the energy (nearly) returns to the initial level for $\hbar\omega=1.5$~eV, below the band gap of \STO{} shown in Fig.~\ref{fig:mSmL-TotalEnergy}. 
The change in the total energy at the maximum of the pulse envelope is larger for photon energies below the gap compared to the absorbing case. As the material is transparent to light with this frequency, the energy and other transient modifications are restored after the pulse.

Our real‑time TDDFT calculations allow us to disentangle the two mechanisms, IFE and OO.
Fig.~\ref{fig:Nexc-mL_IFE-vs-OO} shows the decomposition of the number of excited electrons and the induced orbital magnetization at the in-plane O sites for two pulses discussed in the main manuscript with (a,b)
$\hbar\omega=2.5$~eV for an intensity of $10^{11}$~W\,cm$^{-2}$  and (c,d) $10^{12}$~W\,cm$^{-2}$, as well as (e,f) for $\hbar\omega=1.0$~eV with an intensity of $10^{11}$~W\,cm$^{-2}$. 
The decomposition holds nearly perfectly for the total number of excited carriers, as demonstrated in Fig.~\ref{fig:Nexc-mL_IFE-vs-OO}(a,c,e) and approximately for the orbital magnetization of the in-plane O-$2p$ states, which form the majority of initial states, see Fig.~\ref{fig:Nexc-mL_IFE-vs-OO}(b,d,f). 

For  photon energies below the band gap IFE is the sole contribution to both quantities, as illustrated in Fig.~\ref{fig:Nexc-mL_IFE-vs-OO}(f).
For excitations above the band gap ($\hbar\omega = 2.5$~eV) the two mechanisms compete for an intensity of $10^{11}$~W\,cm$^{-2}$, see Fig.~\ref{fig:Nexc-mL_IFE-vs-OO}(b).  In this regime the OO contribution appears at later times and saturates after the pulse.
With increasing intensity in Fig.~\ref{fig:Nexc-mL_IFE-vs-OO}(d) the IFE becomes stronger during the pulse, whereas the OO
component saturates although the number of persistently excited electrons still increases (see Fig.~\ref{fig:Nexc-mL_IFE-vs-OO}(c)). This indicates that additional non-linear processes may be relevant, which modify  depending on the pulse strength the proportionality between the number of excited carriers and resulting orbital magnetization. In general, this relation will also change with frequency, as the transitions depend on the density of the involved initial and final states.

Figure~\ref{fig:Nexc-mL_IFE-vs-OO_intensity} displays the maximum number of excited carriers (a) and orbital
magnetization of the in‑plane O atoms (b) versus the laser intensity. Below the gap both quantities scale linearly with intensity, reflecting a pure IFE contribution.  Above the gap the linear relationship holds only at low intensities; it breaks down as the intensity increases and OO becomes significant.  For sub‑gap excitation the OO contribution is essentially absent, except at the highest intensities explored.  Conversely, for above‑gap
excitation the OO term rises already at modest intensities and then saturates. For a specific frequency this limits the spin~\cite{zakharchenya1984optical,DAlessandro2020}
or orbital~\cite{Scali2026} polarization that can be achieved, as the excitation depends on the availability of initial and final states. 

\end{document}